\documentclass[aps,pre,floatfix,nofootinbib,showpacs]{revtex4} 
\usepackage{graphicx} \usepackage{amsmath} \usepackage{amssymb} 
\usepackage[sort&compress]{natbib}
\usepackage{graphicx}
\usepackage{amsmath}
\usepackage{amssymb}
\newcommand{\BEQ}{\begin{equation}}
\newcommand{\EEQ}{\end{equation}}
\newcommand{\BEA}{\begin{eqnarray}}
\newcommand{\EEA}{\end{eqnarray}}

\renewcommand{\d}{{\rm d}}

\newcommand{\xb}{{\bf x}}
\newcommand{\yb}{{\bf y}}

\newcommand{\ub}{{\bf u}}
\newcommand{\epsil}{\varepsilon}
\newcommand{\ep}{\varepsilon}

\newcommand{\E}{{\cal E}}

\newcommand{\eps}{\varepsilon}

\newcommand{\om}{\omega}

\newcommand{\half}{\frac{1}{2}}
 
\newcommand{\1}{{\rm S}_{1}} 
\newcommand{\2}{{\rm S}_{2}} 
 
\newcommand{\x}{\hat{x}} 
\newcommand{\p}{\hat{p}} 
 
\newcommand{\hrho}{\hat{\rho}} 
\begin{document} 
\draft
\title
{Brownian Entanglement.}
\date{Today:~\today}

\author{A.E. Allahverdyan$^{1,2)}$, A. Khrennikov$^{3)}$
and Th.M. Nieuwenhuizen$^{1)}$}
\affiliation{$^{1)}$ Institute for Theoretical Physics,
Valckenierstraat 65, 1018 XE Amsterdam, The Netherlands}
\affiliation{$^{2)}$Yerevan Physics Institute,
Alikhanian Brothers Street 2, Yerevan 375036, Armenia}
\affiliation{$^{3)}$International Centre for Mathematical Modeling
in Physics and Cognitive Sciences,\\
MSI, University of V\"axj\"o, S-35195, Sweden}

\begin{abstract} 

We show that for two classical brownian particles there exists an
analog of continuous-variable quantum entanglement: The common
probability distribution of the two coordinates and the corresponding
coarse-grained velocities cannot be prepared via mixing of any
factorized distributions referring to the two particles in
separate. This is possible for particles which interacted in the
past, but do not interact in the present.  Three factors are crucial
for the effect: 1) separation of time-scales of coordinate and momentum
which motivates the definition of coarse-grained velocities; 2) the
resulting uncertainty relations between the coordinate of the brownian
particle and the change of its coarse-grained velocity; 3) the fact
that the coarse-grained velocity, though pertaining to a single
brownian particle, is defined on a common context of two
particles. The brownian entanglement is a consequence of a
coarse-grained description and disappears for a finer resolution of
the brownian motion. We discuss possibilities of its experimental
realizations in examples of macroscopic brownian motion.

\end{abstract}


\maketitle

\section{Introduction.}

There is a long tradition of seeking connections between quantum
mechanics and classical statistical physics. A list of known
examples includes {\it i)} analogies 
between the notion of complementarity in quantum mechanics and statistical
thermodynamics, in particular, between quantum mechanical
uncertainty relations and the energy-temperature  
uncertainty relation in statistical thermodynamics
\cite{bohruncertain,sch,rosenfeld,landau}; {\it ii)} mathematical relations
between Schr\"odinger and Fokker-Planck equations \cite{risken}, which
makes quantum intuition very useful, e.g., for polymers
\cite{gros}; {\it iii)}
attempts to derive Schr\"odinger equations from the classical kinetic
picture \cite{nelson} or from classical stochastic electrodynamics
\cite{timo}.

More generally, both quantum mechanics and statistical physics are
essentially probabilistic theories and already at this level one
expects to find certain analogies between their concepts.  The
progress in this direction has been for a long time plagued by
statements on incompatibility between quantum mechanics and classical
probability theory, which lies in the basis of classical statistical
physics. As an example see Ref.~\cite{feynman} for statements that
quantum interference phenomena contradict to classical probability
theory.  It was, however, noted that such opinions are not warranted
\cite{ball}, and recently it was shown explicitly that several basic
relations of quantum mechanics can be derived from the
classical probability theory provided the contexts of physical
conditions (measurements) are properly taken into account \cite{and}.

Once the basic probabilistic ground of quantum mechanics and classical
statistical physics is recognized to be the same, one wonders whether
some unusual aspects of quantum mechanics, such as entanglement, can
find analogies in classical statistical physics. We see at least two
reasons for seeking such analogies. First, it is going to refresh our
understanding of classical statistical physics, and may imply in
future that advantages offered by quantum mechanics in certain tasks
of information processing and transfer are not unique to quantum
mechanics, and can be looked for in classical areas of physics as well
\cite{and}.  Second, it is useful for understanding quantum mechanics
that certain concepts believed to be purely quantum |that is,
incomprehensible in classical terms| can find natural classical
analogies. In fact the works and ideas mentioned in the above point
{\it i)--iii)} were partially directed toward this goal.  More recent
results along these lines are classical analogies to quantum
entanglement found in optics \cite{spr,italo} and classical
information theory \cite{pop} (secret classical correlations), and a
classical probabilistic model for certain aspects of (multi-time)
quantum measurements \cite{ki}.

The purpose of the present paper is to show that quantum entanglement
can have a natural analogy in the physics of brownian particles
(brownian entanglement).  The reason for the existence of this analogy
can be qualitatively explained as follows. It is known that the
dynamics of a brownian particle can be observed at two levels
\cite{risken}. Within the first, more fundamental level the brownian
particle coupled to a thermal bath at temperature $T$ is described via
definite coordinate $x$ and momentum $p$ and moves under influence of
external potential, friction force and an external random force.  The
latter two forces are generated by the bath. The second, overdamped
regime applies when the characteristic relaxation time of the
coordinate $\tau_x$ is much larger than that of the momentum $\tau_p$,
$\tau_x\gg\tau_p$ (overdamped regime). On times much larger than
$\tau_p$ one is interested in the change of the coordinate and defines
the {\it coarse-grained} velocity as $v=\Delta x/\Delta t$ for
$\tau_x\gg \Delta t\gg \tau_p$.  This definition of $v$ is the only
operationally meaningful one for the (effective) velocity within the
overdamped regime.  It appears that the coarse-grained velocity,
though pertaining to single particles, is defined in the context of
the whole systems of coupled brownian particles. Together with
uncertainty relations between the coordinate and the change of the
coarse-grained velocity |the role of Planck's constant is being played
by the temperature of the bath,| this contextuality feature will be
shown to cause a phenomenon similar to quantum entanglement: The
common probability distribution of the two coordinates and the
corresponding coarse-grained velocities |for two brownian particles
which interacted in the past, but need not interact in the present||
cannot be prepared via mixing of any factorized distributions
referring to the two particles separately.  This brownian entanglement
is a consequence of a coarse-grained description and disappears within
the first (more fundamental) level of description, simply because
entanglement is absent in classical mechanics.

The paper is organized as follows.  In section \ref{qm} we recall the
phenomenon of (continuous-variable) quantum entanglement, focusing
especially on the relations between the entanglement and the
uncertainty relations. Section \ref{III} discusses coarse-grained
velocities and uncertainty relations for the classical brownian
motion.  Next two sections define and study the phenomenon of brownian
entanglement. Sections \ref{artashir} and
\ref{samson} offers detailed comparison between the features of
quantum entanglement and those of its brownian counterpart. In section
\ref{expo} we discuss possibilities of experimental realization of
brownian entanglement. Our conclusions are presented in the last
section. Some technical questions are worked out in Appendix.

\section{Quantum entanglement}
\label{qm}

\subsection{Statistical interpretation.}

This section recalls the phenomenon of entanglement in quantum
mechanics, and especially underlines its connections with the
uncertainty relations. 

Before starting, it is useful to stress that in the present paper
we adhere to the statistical (ensemble) interpretation of
quantum mechanics,
where a quantum `state' is described by a density matrix $\hat\rho$,
and any
state, including a pure state $|\psi\rangle\langle\psi|$, refers to an
ensemble $\E(\hrho)$ of identically prepared systems; see, e.g.,
\cite{blokh,ballentine,newton,balian,home,espagnat,willi}
~\footnote{
The minimal statistical interpretation should of course be
distinguished from various hidden-variable theories and assumptions,
in particular, from the pre-assigned initial value assumptions, where
quantum measurements are viewed as merely revealing pre-existing
values of all observables. Unfortunately, some proponents
of the statistical interpretation were
unclear at this point, a fact that for a while discredited this
interpretation. For a well-balanced discussions on this
and related points, see \cite{espagnat,willi}.}. To put it succinctly:
quantum mechanics makes\footnote{This way of putting the message of
statistical interpretation is adopted from \cite{kruger}.}
\begin{itemize}

\item statistical statements on

\item the results of measurements done

\item on ensembles of identically prepared systems.

\end{itemize}

As was stressed repeatedly
\cite{blokh,ballentine,newton,balian,home,espagnat,willi},
in particular by experimentalists \cite{itano}, 
the experimentally relevant statements of quantum mechanics do not
require more than the minimal statistical interpretation
\footnote{Still, as correctly pointed out in \cite{ocohen,jaynes},
the choice of interpretation can and does influence one's estimates of
importance for various scientific problems.}.
Moreover,
this interpretation deals more successfully with the measurement
problem, as instanced by a recent exactly solvable model
\cite{cw}, and allows to reconcile quantum mechanics with
classical probability theory \cite{and}.  The fact that discussions on
quantum entanglement (and on Bell inequalities and related matters) do
not employ the statistical interpretation is a mere prejudice; see
Refs.
\cite{ballentine,ocohen,kruger,cw} for examples of such discussions.

\subsection{Definition of entanglement.}

Consider a quantum system S consisting of two subsystems $\1$ and $\2$. A
state $\hrho$ of S is called entangled, see e.g. \cite{peres},
(or non-separable) with respect to $\1$ and $\2$, if it {\it cannot} be 
represented as
\BEA
\label{1}
\hrho=\sum_{k=1}^np_k\,\hrho^{(1)}_k\otimes\hrho^{(2)}_k,\qquad
\sum_{k=1}^np_k=1, \qquad p_k\ge 0,
\EEA
where $\hrho^{(1)}_i$ and $\hrho^{(2)}_i$ are arbitrary density matrices
living in the Hilbert spaces of $\1$ and $\2$, respectively, 
$n$ is an integer, and where $\{p_k\}_{k=1}^n$ is a probability distribution. 

According to definition (\ref{1}), a separable quantum state can
always be prepared by means of mixing \footnote{Mixing ensembles
$\E(\hrho_1)$ and $\E(\hrho_2)$ with probabilities $p_1$ and $p_2$,
respectively, means that one throws a dice with probabilities of
outcomes equal to $p_1$ and $p_2$, and depending on the outcome one
picks up a system from $\E(\hrho_1)$ or $\E(\hrho_2)$, keeping no
information on where the system came from. Alternatively, one can join
together $Np_1$ systems from $\E(\hrho_1)$ and $Np_2$ systems from
$\E(\hrho_2)$ ($N\gg 1$), so that no information information is kept
on where a single system came from. } 
non-correlated states $\hrho^{(1)}_k\otimes\hrho^{(2)}_k$
of the two subsystems. For a pure state $\hrho
=|\psi\rangle\langle\psi|$ we return to the more known definition of
entanglement: $|\psi\rangle$ cannot be represented as
$|\psi\rangle=|\psi\rangle_1\otimes|\psi\rangle_2$. For this
particular case the absence of entanglement implies the absence of any
correlation.  In contrast, for the more general situation given by
(\ref{1}), a non-entangled state can still posses certain (classical)
correlations, since the totally uncorrelated situations will be given
as $\hrho=\hrho^{(1)}\otimes\hrho^{(2)}$.

Two important features of quantum entanglement should be noted.  First,
it can be tested only through measuring some correlations between the
subsystems $\1$ and $\2$. Observables pertaining to $\1$ or to $\2$ alone
are obviously insensitive to entanglement.  Second, if the state
(\ref{1}) is prepared by two different observers 1 and 2, then this
preparation is seen to involve correlated actions of them, and thus
the corresponding observers have to communicate classically.

\subsection{Classical systems.}
\label{claa}

In classics the representation (\ref{1}) is apparently
always possible. Indeed,
let us have a probability distribution $P(x_1,x_2)$ of two classical
systems $\1$ and $\2$ represented by
random variables $x_1$ and $x_2$. Assume for simplicity that
$x_1$ can take values $a_1,..,a_n$, 
while $x_2$ can take values $b_1,..,b_n$.
Then $P(x_1,x_2)$ can be written as
\BEA
P(x_1,x_2)=\sum_{\alpha=a_1,...,a_n}~\sum_{\beta=b_1,...,b_n}
P(\alpha,\beta)\,
\delta_{\alpha x_1}\, \delta_{x_2\beta},
\label{111}
\EEA
where $\delta_{\alpha\beta}=1$ if $\alpha=\beta$, and
$\delta_{\alpha\beta}=0$ otherwise. {\it Provided} that
$\delta_{\alpha x_1}$ and $\delta_{x_2\beta}$ are two legitimate
probability distributions belonging to $\1$ and $\2$, that is,
provided there are no mechanisms prohibiting the realization of
$\delta_{\alpha x_1}$ and $\delta_{x_2\beta}$ as physically acceptable
distributions usable in the actual preparation, the representation
(\ref{1}) is realized for the classical situation: there is no
entanglement.

In fact the quantum situation goes to the classical one right in (\ref{1})
if we assume that the involved density matrices are always
diagonal. So it is the presence of non-diagonal elements of the given
density matrix that makes the situations different.  

\subsection{Semiclassical systems.}

We shall now localize the cause of quantum entanglement for systems
which behave (semi)classically in several other respects.  This will
help us to understand the way of searching for entanglement in
non-quantum situations.

Let $\1$ and $\2$ be two quantum (non-interacting) harmonic
oscillators, and the overall density matrix (in the coordinate
representation) $\rho(x_1,x'_1;x_2,x'_2)$ be a gaussian function of
its variables.  The corresponding coordinate and momenta operators are
denoted by $\x_1,\,\p_1$ and $\x_2,\,\p_2$, respectively. As
well-known, gaussian states are conveniently dealt with help of Wigner
functions $W(x,p)$, which are functions of $x$ and $p$
\footnote{For good reviews on the
properties of Wigner function see
\cite{wignerreview}.}. The
Wigner function is equivalent to the density matrix, that is, it
allows to calculate all averages, but in contrast to the density
matrix it has several properties expected for the common probability
distribution of the coordinate and momentum. In particular, the
analysis via Wigner functions will provide us below with a richer
intuition on the relations between entanglement and uncertainty relations;
thus, between entanglement and non-commutativity.
 
Since the overall Hamiltonian of $\1$ and $\2$ is assumed to be
harmonic, the initially gaussian state remains gaussian for all times
and the corresponding Wigner function will be positive and thus will
{\it partially} admit a classical interpretation in terms of common
probability distribution of the coordinate and momentum
\cite{wignerreview}. A related
fact is that for two harmonic oscillators the Ehrenfest equations of
motion can be recast into a classical form
\cite{wignerreview}. So many aspects of this system can be accounted
for in classical terms.

In order to see why in spite of these classical features the system
of two oscillators can be entangled, note that for the Wigner functions one
can rewrite the condition (\ref{1}) as
\BEA
W(x_1,p_1;x_2,p_2)=\int \d \lambda \,{\cal P}(\lambda)\,
W_1(x_1,p_1|\lambda)\,W_2(x_2,p_2|\lambda),
\qquad {\cal P}(\lambda)\geq 0,\qquad
\int\d \lambda\,{\cal P}(\lambda)=1.
\EEA
where $W_1$ and $W_2$ are separate Wigner functions for $\1$ and
$\2$ respectively, and ${\cal P}(\lambda)$ is some probability 
distribution. If $\1$ and $\2$ were classical oscillators,
then instead of Wigner functions we would have distribution functions,
and one can always
write down the analog of (\ref{111}) for the common distribution function
$P(x_1,p_1;x_2,p_2)$ and 
$\lambda=(\alpha_1\,\beta_1,\alpha_2,\beta_2)$:
\BEA
\label{222}
P(x_1,p_1;x_2,p_2)=\int \d \alpha_1\,\d \beta_1\,\d \alpha_2\,
\d \beta_2\,P(\alpha_1,\beta_1;\alpha_2,\beta_2)\,
\delta(\alpha_1-x_1)\,\delta(\beta_1-p_1)\,
\delta(\alpha_2-x_2)\,\delta(\beta_2-p_2).
\EEA
In the classical situation this means that there is no
entanglement. The same formula (\ref{222}) can formally be written
down also for the positive Wigner functions $W(x_1,p_1;x_2,p_2)$.
However, in the quantum situation
$\delta(\alpha_1-x_1)\,\delta(\beta_1-p_1)$ and
$\delta(\alpha_2-x_2)\,\delta(\beta_2-p_2)$ are {\it not} legitimate
Wigner functions, since they prescribe definite values to both
coordinate
and momentum and thus
do not respect the
uncertainty relations.

We conclude that the uncertainty relations are necessary |but
not sufficient| for the existence of entanglement in
semiclassical systems.

\subsection{A simple sufficient condition for quantum entanglement}

The definition of the entanglement as given by Eq.~(\ref{1}) is not
practical (except for a pure density matrix $\hrho$, when no
entanglement means no correlations). It is therefore useful to have
certain sufficient conditions for the presence of entanglement which
will be easy to handle in applications and which will have a
transparent physical meaning.

We choose the units in such a way that the coordinate and the
momentum have the same dimension as $\sqrt{\hbar}$. For a harmonic
oscillator with mass $m$ and frequency $\omega$ it will suffice to
make the following canonical transformation: $\x\to\sqrt{m\omega}\,\x$,
$\p\to\p/\sqrt{m\omega}$. 

For a two-particle system S with coordinate and momenta operators
$\hat{x}_1$, $\hat{x}_2$ and $\hat{p}_1$, $\hat{p}_2$, respectively, one
can propose the following sufficient condition for the entanglement
\cite{hoffmann}. 

Let us first note that the standard uncertainty relation 
\begin{gather}
\label{dub1}
\langle \Delta\hat{x}^2\rangle\,\langle \Delta\hat{p}^2\rangle
\geq \frac{\hbar^2}{4},\\
\Delta \x\equiv\x-\langle\x\rangle,\qquad
\Delta \p\equiv\p-\langle\p\rangle,
\end{gather}
implies
\BEA
\label{dalal}
&&\langle \Delta\hat{x}^2\rangle+\langle \Delta\hat{p}^2\rangle
\geq\langle \Delta\hat{x}^2\rangle+\frac{\hbar^2}
{4~\langle \Delta\hat{x}^2\rangle},
\EEA
and then one gets via minimizing the RHS of (\ref{dalal}) over $\langle
\Delta\hat{x}^2\rangle$ (which produces
$\langle\Delta\hat{x}^2\rangle\to\hbar/2$, to be
put in the RHS of (\ref{dalal})):
\BEA
\langle \Delta\hat{x}^2\rangle+\langle \Delta\hat{p}^2\rangle
\geq \hbar.
\label{dub2}
\EEA

Now assume that the two-particle system is described by a factorized
density matrix:
\BEA
\hrho=\hrho^{(1)}\otimes\hrho^{(2)}.
\label{gambar}
\EEA
Then due to (\ref{dub2}) one can write:
\BEA
\label{dub3}
\langle (\Delta\hat{x}_1-\Delta\hat{x}_2)^2\rangle+
\langle (\Delta\hat{p}_1+\Delta\hat{p}_2)^2\rangle
=\langle \Delta\hat{x}_1^2\rangle+\langle \Delta\hat{p}_1^2\rangle
+\langle \Delta\hat{x}_2^2\rangle+\langle \Delta\hat{p}_2^2\rangle
\geq 2\hbar,
\EEA
just because for non-correlated systems the corresponding
variances add up.  If the overall system is in a separable state, then
(\ref{dub3}) is even strengthened, since the variance of {\it any} observable
$\langle\Delta\hat{A}^2\rangle$
increases under mixing, i.e. under a transformation:
\BEA
\{p_k,\hrho_k\}
\to \sum_kp_k\hrho_k,\qquad p_k\geq 0, \qquad\sum_kp_k=1,
\EEA
where $\hrho_k$ are normalized (${\rm tr}\,\hrho_k=1$)
density matrices. Indeed,
\BEA
\langle\Delta\hat{A}^2\rangle=\sum_kp_k{\rm tr}\,[\,
\hrho_k\,(\hat{A}-\langle\hat{A}\rangle)^2 \,]&&=
\sum_kp_k\left\{{\rm tr}\,[\,\hrho_k\,\hat{A}^2\,]
-{\rm tr}\,[\,\hrho_k\,\hat{A}\,]^2\right\}+
\sum_kp_k\left\{{\rm tr}\,[\,\hrho_k\,\hat{A}\,]
-\langle\hat{A}\rangle\right\}^2\nonumber\\
&&=\sum_kp_k\langle\Delta\hat{A}^2\rangle_k
+\sum_kp_k\left\{{\rm tr}\,[\,\hrho_k\,\hat{A}\,]
-\langle\hat{A}\rangle\right\}^2\geq
\sum_kp_k\langle\Delta\hat{A}^2\rangle_k.
\label{lulu}
\EEA
Employing (\ref{lulu}) for 
$\hat{A}=\Delta\hat{x}_1-\Delta\hat{x}_2$ and
for $\hat{A}=\Delta\hat{p}_1-\Delta\hat{p}_2$, we get that
though (\ref{dub3}) was obtained for 
the factorized state (\ref{gambar}), it remains valid 
for an arbitrary non-entangled state 
$\sum_kp_k\hrho^{(1)}_k\otimes\hrho^{(2)}_k$. Thus the violation
\BEA
\langle (\Delta\hat{x}_1-\Delta\hat{x}_2)^2\rangle+
\langle (\Delta\hat{p}_1+\Delta\hat{p}_2)^2\rangle
\leq 2\hbar
\label{buenosaires}
\label{klm}
\EEA
of (\ref{dub3}) 
is a sufficient condition for entanglement. 

Eq.~(\ref{buenosaires}) has a transparent physical meaning:

\begin{itemize}

\item
entanglement is present, if fluctuations do not sum up additively,
i.e., if the changes $\Delta\hat{x}_1$ and $\Delta\hat{x}_2$ of the
coordinates tend to correlate with each other, while those of the
momenta tend to anticorrelate. 
\end{itemize}

Alternative sufficient conditions for entanglement 
can be built based on $\langle
(\Delta\hat{x}_1+\Delta\hat{x}_2)^2\rangle+ \langle
(\Delta\hat{p}_1-\Delta\hat{p}_2)^2\rangle$ or $\langle
(\Delta\hat{x}_1+\Delta\hat{x}_2)^2\rangle+ \langle
(\Delta\hat{p}_1+\Delta\hat{p}_2)^2\rangle$, etc. Various conditions
obtained in this way are obviously not equivalent to each other. 
An obvious way to strengthen it is to demand that
\BEA
\label{lapas}
\langle (\Delta\hat{x}_1+\epsilon
\Delta\hat{x}_2)^2\rangle+ \langle
(\Delta\hat{p}_1+\zeta\Delta\hat{p}_2)^2\rangle\leq 2\hbar,
\EEA
at least
for one of four independent choices $\epsilon=\pm 1$, $\zeta= \pm 1$.

For some special states (e.g. gaussian states) there exist
in literature necessary and sufficient conditions for entanglement
\cite{wolf}.  However, they are technically involved and do not have
such a straightforward physical meaning as (\ref{buenosaires}).
For our purposes conditions (\ref{buenosaires}, \ref{lapas})
are sufficient.

\subsection{Operational issues.}
\label{oper}

Let us finally recall how the condition (\ref{buenosaires})
is checked operationally. To this end, we re-write this inequality
as
\BEA
\label{ap1}
&&\langle \Delta\hat{x}_1^2\rangle +\langle\Delta\hat{x}_2^2\rangle+
\langle \Delta\hat{p}_1^2\rangle+\langle\Delta\hat{p}_2^2\rangle\\
&&-2\langle \Delta\hat{x}_1\,\Delta\hat{x}_2\rangle+
2\langle \Delta\hat{p}_1\,\Delta\hat{p}_2\rangle
\label{ap2}\\
&&\leq 2\hbar.\nonumber
\EEA
One now needs two ensembles of systems S, each one
consisting of identically prepared
correlated sub-systems $\1$ and $\2$. Assuming that $\1$
and $\2$ are in possession of observers 1 and 2, respectively, the
observer 1 measures on the first (second) ensemble $\hat{x}_1$ ($\hat{p}_1$),
while the observer 2
measures on the first (second) ensemble $\hat{x}_2$ ($\hat{p}_2$).
Note that the above two ensembles are necessary, since 
$\hat{x}_1$ and $\hat{p}_1$ 
(respectively, $\hat{x}_2$ and $\hat{p}_2$) do not commute:
$[\hat{x}_k,\hat{p}_l]=i\hbar\delta_{kl}$, with $\delta_{kl}$
being Kronecker's delta, and thus cannot be measured simultaneously.  
Now the quantities in (\ref{ap1}) can be
estimated by each observer separately, while for the quantities in
(\ref{ap2}) the observers need to put the results of their
measurements together (or to communicate in any other classical way)
and to count the coinciding events.

\section{Coarse-grained velocities and
uncertainty relations for brownian particles.}
\label{III}

Consider $N$ identical brownian particles with coordinates
$\xb=(x_1,...,x_N)$ and mass $m$ interacting with $N$ independent
thermal baths at temperatures $T_i$ and subjected to a potential
$U(x_1,...,x_N)$.  The overdamped limit is defined by the following
two conditions \cite{risken} (for a more detailed discussion see
Appendix):

\begin{itemize} 

\item The characteristic relaxation time 
\footnote{The characteristic relaxation time 
$\tau_\theta$ of a random time-dependent variable $\theta$
(e.g. coordinate or momentum) is an {\it ensemble notion} and is
defined as the time necessary for the conditional distribution of
$\theta$ to become memoryless: $P(\theta,t'+\tau_\theta |\theta',t')
=P(\theta,t'+\tau_\theta )$.  For the system of brownian particles
sufficiently strongly coupled to their thermal baths, the
arbitrarinesses of the above definition (e.g., the precise choice of
$\theta'$ and $t'$) do not change the physical content of the
definition \cite{risken}. } of the (real) momenta $m\dot{x}_i$ is much
smaller than the one of the coordinates.

\item One is interested in times which are much larger than the
relaxation time of the momenta, but which can be much smaller than
or comparable
to the relaxation time of the coordinates.

\end{itemize}

Under these conditions the dynamics of the system is described
by the following Langevin equations \cite{risken}:
\BEA
\label{langevin}
\dot{x}_i=f_i(\xb)+\eta_i(t),\qquad f_i(\xb)=-\partial_{x_i} U(\xb),
\qquad \langle\eta_i(t)\,\eta_j(t')\rangle=2\,T_i\,\delta_{ij}\,\delta(t-t'),
\EEA
where for our convenience 
both the mass $m$ and the damping constant (coupling constant of
the particles to the bath) are in the main text taken equal to one
(they are recorded in the Appendix)
\footnote{Eq.~(\ref{langevin}) can be obtained from
the complete Langevin equations of motion: [with the same definition
of $\eta_i(t)$ as in (\ref{langevin})]
$m\ddot{x}_i+\dot{x}_i=f_i(\xb)+\eta_i(t)$, by disregarding the first
term $m\ddot{x}_i$ corresponding to acceleration. One might make this
heuristically by requiring that the friction force $\dot{x}_i$
dominates, and that the second-derivative $\ddot{x}_i$ is small for
long times. More rigorous derivation of this classical problem is
presented in \cite{risken}, and is recalled in Appendix for a simple
model.  In their turn the above complete Langevin equations can be
rigorously derived from the Newton equations of motion for the
brownian particles and the baths. Here are the basic conditions for
this derivation. 1) The thermodynamical (macroscopic) limit for the
baths. 2) A reasonable model for the particle-bath interaction. 3) The
initially equilibrium state of the baths, described by the
corresponding Gibbs distributions at temperature $T_i$ (this is how
the random noises $\eta_i(t)$ come into existence).  The details of
the derivation can be looked up, e.g., in \cite{cl,gardiner}.  }.

The conditional probability $P(\xb,t|\xb ',t')$ is known to satisfy the 
following Fokker-Planck equation \cite{risken}:
\BEA
\label{fokkerplanck}
\label{d1}
\partial_t P(\xb,t|\xb ',t')
=-\sum_i\partial_{x_i}\,[\, f_i(\xb)\,P(\xb,t|\xb ',t') \,]+
\sum_iT_i\,\partial^2_{x_ix_i}P(\xb,t|\xb ',t'),\qquad
t\geq t'.
\EEA
Eq.~(\ref{d1}) is associated with the following initial condition:
\BEA
\label{d2}
P(\xb,t|\xb ',t)=\delta(\xb-\xb ')\equiv\prod_{i=1}^N\delta(x_i-x'_i),
\EEA
as follows from the very definition of the conditional probability.

Consider an ensemble $\Sigma(\xb,t)$ of all realizations of the whole
$N$-particle system which at time $t$ have a coordinate vector
$\xb$. Such an ensemble can be selected out of all possible realizations
by measuring $\xb=(x_1,...,x_N)$.  For this ensemble the average coarse-grained
velocity for the brownian particle with index $j$ might naively be
defined as:
\BEA
v_{j}(\xb,t)={\rm lim}_{\ep\to 0}
\,\int\d \yb\,\frac{y_j-x_j}{\ep}\,P(\yb,t+\ep|\xb,t).
\EEA
However, the absence of regular trajectories
enforces us to define different velocities for different directions of
time \cite{nelson}:
\BEA
\label{dish1}
v_{+,j}(\xb,t)={\rm lim}_{\ep\to +0}
\,\int\d y_j\,\frac{y_j-x_j}{\ep}\,P(y_j,t+\ep|\xb,t),\\
v_{-,j}(\xb,t)={\rm lim}_{\ep\to +0}
\,\int\d y_j\,\frac{x_j-y_j}{\ep}\,P(y_j,t-\ep|\xb,t).
\label{dish2}
\EEA

The physical meaning of these expressions can be explained as follows.

\begin{enumerate}

\item As seen from the definitions, $v_{+,j}(\xb,t)$ is the average 
velocity to move anywhere starting from $(\xb,t)$, whereas
$v_{-,j}(\xb,t)$ is the average velocity to come from anywhere and to
arrive at $\xb$ at the moment $t$.

Since these velocities are defined already in the overdamped limit,
$\ep$ is assumed to be much larger than the characteristic relaxation
time of the (real) momentum which is small in the overdamped limit.
Therefore, we call (\ref{dish1}, \ref{dish2}) coarse-grained
velocities. It is known that for the overdamped brownian motion almost
all trajectories are not smooth. This is connected to the chaotic
influences of the bath(s) which randomize the real momenta on much
smaller times, and this is also the reason for $v_{+,j}(\xb,t)\not
=v_{-,j}(\xb,t)$. The difference $v_{+,j}(\xb,t)- v_{-,j}(\xb,t)$
thus characterizes the degree of the above non-smoothness.

Recall that would one take $\ep$ much smaller than the
characteristic relaxation time of the momentum |which will amount to
applying definitions (\ref{dish1}) and (\ref{dish2}) to a smoother
trajectory| then $v_{+,j}(\xb,t)$ and $v_{-,j}(\xb,t)$ will be equal
to each other and equal to the average momentum.  These points are
discussed in detail in Appendix.

\item Here is an operational procedure for measuring 
$v_{+,j}(\xb,t)$ and $v_{-,j}(\xb,t)$.
Consider the overall ensemble of the brownian particles.
The single members of this ensemble are $N$ coupled brownian
particles. For each such single member one measures

{\it i)} the coordinate of the brownian particle with index $j$
at the moment $t-\varepsilon$;

{\it ii)} the coordinates of {\it all} brownian particles 
at the moment $t$;

{\it iii)} the coordinate of the same brownian particle with index $j$
at the moment $t+\varepsilon$.

Repeating these points many times on various members of the above
overall ensemble, ignoring all the results from {\it i)} and {\it
iii)}, and selecting events from the second step, we reconstruct the
ensemble $\Sigma(\xb,t)$. Employing the results from {\it i)} and {\it
iii)} and conditioning repeatedly upon a single member from
$\Sigma(\xb,t)$, we estimate the conditional probabilities
$P(y_j,t\pm\ep|\xb,t)$ and then finally calculate $v_{+,j}(\xb,t)$ and
$v_{-,j}(\xb,t)$ via Eqs.~(\ref{dish1}, \ref{dish2}).

\item The above ensemble $\Sigma(\xb,t)$ is different
from an ensemble $\Sigma_j(x_j,t)$ obtained by measuring $x_j$ (at
time $t$) {\it irrespective} of other coordinates. Indeed,
$\Sigma(\xb,t)$ is defined globally, and if there are different
observers in possession of each brownian particle, they have to
communicate to each other in order to be able to construct
$\Sigma(\xb,t)$. In contrast, $\Sigma_j(x_j,t)$ is defined exclusively
with respect to the brownian particle with index $j$.  This point will
be discussed in more details later on in section
\ref{artashir}.

\end{enumerate}

The calculation of $v_{+,i}(x,t)$ and $v_{-,i}(x,t)$
is straightforward upon using the following three things: relation
\BEA
P(\xb,t+\ep|\yb,t)=\delta(\xb-\yb)+\ep\sum_i\left[
-f_i(\yb)\partial_{x_i}\,\delta(\xb-\yb)+
T_i\,\partial^2_{x_ix_i}\delta(\xb-\yb)
\right],
\EEA
which follows from (\ref{d1}, \ref{d2}), Bayes formula, and partial
integration assuming natural boundary conditions at infinity: $P(x,t)\to
0$ if $x_j\to \pm\infty$. 
Starting from the definitions (\ref{dish1}, \ref{dish2}), we obtain
\BEA
\label{shun}
v_{+,j}(\xb,t)&=&
{\rm lim}_{\ep\to +0}
\,\int\d \yb\,\frac{y_j-x_j}{\ep}\,P(\yb,t+\ep|\xb,t)\nonumber\\
&=&\,\int\d \yb\,(y_j-x_j)\sum_i
\left[\,
- f_i(\xb)\partial_{y_i}\delta(\xb-\yb) 
+T_i\partial^2_{y_iy_i}\delta(\xb-\yb) \,\right]\nonumber\\
&=&f_j(\xb),
\EEA
\BEA
v_{-,j}(\xb,t)&=&{\rm lim}_{\ep\to +0}
\,\int\d \yb\,\frac{x_j-y_j}{\ep}\,P(\yb,t-\ep|\xb,t)\nonumber\\
&=&{\rm lim}_{\ep\to +0}
\,\int\d \yb\,\frac{x_j-y_j}{\ep}\,P(\xb,t|\yb,t-\ep)\,
\frac{P(\yb,t-\ep)}{P(\xb,t)}\nonumber\\
&=&\int\d \yb\,(x_j-y_j)
\,\frac{P(\yb,t)}{P(\xb,t)}
\sum_i\left[-f(\yb)\,\partial_{x_i}\delta(\xb-\yb)
+T_i\partial_{x_ix_i}\delta(\xb-\yb)\right]\,
\nonumber\\
&=&f_j(\xb)-2T_j\,\partial _{x_j}\ln P(\xb,t).
\label{katu}
\EEA

The difference between the coarse-grained
velocities $v_{+,j}(\xb,t)$ and $v_{-,j}(\xb,t)$ is 
\BEA
\label{babek}
u_j(\xb,t)=\frac{v_{-,j}(\xb,t)
-v_{+,j}(\xb,t)}{2}=-T_j\partial_{x_j}\ln P(\xb,t).
\EEA
Recall that this quantity is non-zero due to the action of the thermal
bath (the factor $T_j$ in (\ref{babek})), and due to the fact that at
the coarse-grained level of description almost all trajectories of the
brownian particles are not smooth. Sometimes $u_j(\xb,t)$ is referred
to as osmotic velocity of the brownian particle with index $j$
\cite{nelson}. We shall use this word as a useful short-hand for the
more precise term ``change of the coarse-grained velocity''.

Note that once the interactions between the brownian particles are
absent: $U(\xb)=\sum_kU(x_k)$ |and consequently correlations are absent
as well, $P(\xb,t)=\prod_kP(x_k,t)$, if they were absent initially| the
coarse-grained velocities $v_{\pm,j}$ depend only the corresponding
coordinate: $v_{\pm,j}(\xb,t)=v_{\pm,j}(x_j,t)$.

The regular (or mechanical) counterpart of the coarse-grained velocity
for the brownian particle with unit mass and index $j$ can be
naturally associated with the newtonian force $f_j(\xb)$ as defined by
(\ref{langevin}). When the bath(s) are absent, this is the only
contribution to the coarse-grained velocity.  The fact that velocity
appears to be proportional to force reminds us that we are in the
overdamped regime of description, where, in particular, friction
dominates over acceleration.

\subsection{Uncertainty relations and their interpretation.}
\label{ma}

Uncertainty relations exist not only in quantum mechanics but also in
physics of overdamped brownian motion \cite{ur}. As the very subject
of statistical physics, they arise out of ignorance reasons, or more
precisely due to separation of time-scales: though the real classical
particles involved in the classical brownian motion certainly do have
sharply defined coordinates and momenta at the microscopical level, at
the coarse-grained (overdamped) level of description a brownian
particle does not have a well-defined trajectory, and cannot
posses sharply defined coordinate and coarse-grained velocity, as 
seen below.

We saw above that the coarse-grained velocities $v_{+,j}(\xb,t)$ and
$v_{-,j}(\xb,t)$ are defined with respect to the ensemble
$\Sigma(\xb,t)$ of all brownian particles which at the moment $t$ pass
via coordinate vector $\xb$. On the other hand, $\Sigma(\xb,t)$ is by
itself a subensemble embedded with probability (weight) $P(\xb,t)$ into
the ensemble of all realizations of the random coordinate vector $\xb$
at the time $t$. Thus, both $v_{+,j}(\xb,t)$ and $v_{-,j}(\xb,t)$, as
well as the osmotic velocity $u_j(\xb,t)$ are random quantities as
functions of of the random configuration of $N$ brownian particles ${\bf
x}=(x_1,\cdots, x_N)$.  In other words, this randomness enters via the
context $(\xb,t)$ which was chosen to define the ensemble
$\Sigma(\xb,t)$. 

More specifically, let us focus on the common distribution function 
$P(\xb,\ub;t)$ of
\BEA
\xb=(x_1,...,x_N)\quad {\rm and}\quad \ub=(u_1,...,u_N),
\EEA
\BEA
P(\xb,\ub;t)=P(\xb,t)\prod_{j=1}^N\delta(\,u_j-u_j(\xb,t)\,),
\label{darder}
\EEA
where $u_j(\xb,t)$ is defined by (\ref{babek}).

One has after averaging over $P(\xb,\ub;t)$
\BEA
\label{toto}
\langle u_j\rangle=
\int \d u_j\,\d \xb\, u_j\,P(\xb,t)\,
\delta(\,u_j-u_j(\xb,t)\,)=-T_j\int \d \xb\,
\partial_{x_j}P(\xb,t)
=0,
\EEA
\BEA
\label{gogo}
\langle\,(\,x_k-\langle x_k\rangle\,)\,(\,u_j-\langle u_j\rangle\,)\rangle
=\int \d \xb\, u_j(\xb,t)\,P(\xb,t)\,(x_k-\langle x_k\rangle)
=-T_j\int \d \xb\,(x_k-\langle x_k\rangle)\,\partial_{x_j}P(\xb,t)
=T_j\delta_{kj}.
\EEA
The fact of $\langle u_j\rangle=0$ is natural, since $u_j$ is the
(coarse-grained) velocity difference generated due to the interaction
to the bath.  It is also seen from (\ref{gogo}) that $u_j$ of the
corresponding brownian particle correlates only with its own
coordinate. This is related to the independence of the thermal baths
which act on different brownian particles.

Applying to (\ref{gogo}) the standard Cauchy-Schwartz inequality one
deduces:
\BEA
\langle\,(\,x_k-\langle x_k\rangle\,)^2\,\rangle\,
\langle\,(\,u_j-\langle u_j\rangle\,)^2\,\rangle\,\geq
|\langle\,(\,x_k-\langle x_k\rangle\,)\,(\,u_j-\langle u_j\rangle\,)\rangle|^2
=T^2_j\delta_{kj},
\label{sq}
\EEA 
which implies an uncertainty relation 
between the uncertainty of the coordinate 
and that of the osmotic velocity 
\cite{ur}.

Obviously, the regular counterpart $f_j(x,t)$ of the coarse-grained velocity
does not (and should not) have any uncertainty relation with the coordinate.

\subsection{Discussion.}

Let us discuss in more detail the difference between the physical
meaning of the brownian uncertainty relation (\ref{sq}) as compared to the
standard quantum mechanical uncertainty relation (\ref{dub1}). The
latter refers to two identical quantum ensembles $\E(\,\hrho(t)\,)$
such that on the first (second) ensemble one measures $\hat{x}$
($\hat{p}$), and then combines into the single relation (\ref{dub1})
the obtained dispersions. Note that the frequent interpretation of
(\ref{dub1}) as referring to disturbance of the momentum when measuring
the coordinate (and {\it vice versa}) is in general incorrect
\cite{willi,ozawa}, since, in particular, the quantities $\langle
\Delta\hat{x}^2\rangle$ and $\langle
\Delta\hat{p}^2\rangle$ clearly refer to different measurements done on
different ensembles, rather than to a simultaneous measurement of the
coordinate and momentum done on a single ensemble. More refined arguments
and explicit counterexamples for this point can be found in 
\cite{willi,ozawa}.

In contrast to this, the quantities entering into Eq.~(\ref{sq}) refer
to consecutive measurements done on the same ensemble.  The involved
coarse-grained velocity is operationally well-defined and characterizes
the change of the coordinate.  This is contrasting to the situation with
the quantum mechanical observable of the momentum which (in the
Heisenberg representation ) in general does {\it not fully} characterize the
change of the coordinate for any initial state; see Footnote \ref{get}
for more details.

\section{Brownian entanglement.}

Consider now two brownian particles.  An analog of entanglement can be
introduced in the following way.  Call the state of two particles
non-entangled (separable) if the common distribution function
(\ref{darder}) of the coordinates and the osmotic velocities can be
prepared by mixing non-correlated distributions
\footnote{We consider only the common distribution function
of the coordinates and the osmotic velocities, and not, e.g.,  the
common distribution of the coordinates and forward velocities
$v_{+,j}(\xb,t)$, simply because due to the brownian uncertainty
relation (\ref{gogo}) only the former can lead to a non-trivial
definition of entanglement.  Recall in this context that the quantum
uncertainty relations are necessary for the existence of the quantum
entanglement.}
\BEA
P(x_1,u_1,x_2,u_2)=\int \d \lambda\,
{\cal P}(\lambda)\,
P_1(x_1,u_1|\lambda)\,P_2(x_2,u_2|\lambda),
\qquad {\cal P}(\lambda)\geq 0,\qquad
\int\d \lambda\,{\cal P}(\lambda)=1.
\label{tartar}
\EEA

We shall naturally require that the separate distributions
$P_1(x_1,u_1|\lambda)$ and $P_2(x_2,u_2|\lambda)$ ensure the basic
properties (\ref{toto}, \ref{gogo}) of the osmotic velocity $u_j$. For
the rest they can be completely arbitrary.  The requirements
(\ref{toto}, \ref{gogo}) make the basic difference as compared to the
(naive) classical discussion in section \ref{claa}, where any
distribution was admissible.  More detailed discussion on the physical
meaning of (\ref{tartar}), and, in particular, on its similarities and
differences with the definition (\ref{1}) of quantum entanglement, is
postponed till section \ref{samson}.

For simplicity we choose units such that $x$ and $u$ have the same
dimension as $\sqrt{T}$ (e.g., we measure time in units of square root
of the damping constant; see Appendix). We take also $T_1=T_2=T$ again
for simplicity. In the same way as we derived (\ref{dub1},
\ref{buenosaires}), one can write the brownian uncertainty relation
(\ref{sq}) as
\BEA
\langle\,\Delta u_1^2\,\rangle+\langle\,\Delta x_1^2\,\rangle\geq 2T,
\qquad
\langle\,\Delta u_2^2\,\rangle+\langle\,\Delta x_2^2\,\rangle\geq 2T,
\EEA
and proceed to derive
\BEA
\label{dodo}
\langle\,(\,\Delta u_1+\Delta u_2\,)^2\,\rangle+
\langle\,(\,\Delta x_1-\Delta x_2\,)^2\,\rangle< 4T,
\EEA
as a sufficient condition for the entanglement. More general relations 
(\ref{lapas}) can also be obviously
transferred to the brownian situation:
\BEA
\label{dodo1}
\langle\,(\,\Delta u_1+\zeta \Delta u_2\,)^2\,\rangle+
\langle\,(\,\Delta x_1+\epsilon \Delta x_2\,)^2\,\rangle< 4T,
\EEA
where $\zeta$ and $\epsilon$ can independently assume values
$\pm 1$.

\subsection{Gaussian state of two interacting brownian particles.}
\label{utkonos}

It will be useful to work out a simple model for two coupled brownian particle,
where the above general concepts can be visualized and studied in detail.

Consider now two harmonically interacting brownian particles
with an overall potential energy
\BEA
U(x_1,x_2)=\frac{ax^2_1}{2}+\frac{ax^2_2}{2}+gx_1x_2,
\label{po}
\EEA
where $a>0$ and $g$ characterizes the interaction between the
particles \footnote{Note that the positivity of 
$U(x_1,x_2)$ is implied by $g^2<a^2$.}. 
The particles interact with independent bath at the same temperature
$T$.

Fokker-Planck equation (\ref{d1})
for the dynamics of these brownian particles
can be solved directly and the outcome is known to be 
given by the following two-dimensional gaussian
(provided the initial distribution was gaussian):
\BEA
\label{zarzand}
P(x_1,x_2;t)
=&&\frac{\sqrt{{\rm det}\,C}}{2\pi}\,
\exp\left[-\half \sum_{i,j=1}^2C_{i,j}x_ix_j
\right],\nonumber\\
C^{-1}=&&
\left(\begin{array}{rr}
\langle x_1^2(t)\rangle~~&~~\langle x_1(t)x_2(t)\rangle \\
\\
\langle x_1(t)x_2(t)\rangle&~~\langle x_2^2(t)\rangle~~ \\
\end{array}\right)\equiv 
\left(\begin{array}{rr}
\sigma_{11}(t)~~&~~\sigma_{12}(t) \\
\\
\sigma_{12}(t)~~&~~\sigma_{22}(t) \\
\end{array}\right),
\EEA
where for simplicity we assumed that 
\BEA
\langle x_1(t)\rangle=\langle x_2(t)\rangle=0,
\EEA
for all times. It is also useful to note the marginal distributions
of $P(x_1,x_2;t)$:
\BEA
\label{taratuj}
P_j(x_j;t)=\exp\left[
-\frac{x_j^2}{2\sigma_{jj}(t)}
\right],\qquad j=1,2.
\EEA

The easiest way to obtain $\sigma_{11}(t)$, $\sigma_{22}(t)$ and
$\sigma_{12}(t)$ is to look directly at the Langevin equations
(\ref{langevin}) with the potential (\ref{po}):
\BEA
\label{lacot}
\dot{x}_1(t)+ax_1(t)+gx_2(t)=\eta_1(t),
\qquad
\dot{x}_2(t)+ax_2(t)+gx_1(t)=\eta_2(t),
\EEA
which can be more conveniently re-written in terms of the relative
coordinate $r_-=(x_1-x_2)/2$ and the center of mass coordinate 
$r_+=(x_1+x_2)/2$:
\BEA
\dot{r}_{\pm}(t)=-(a\pm g)r_\pm(t)+\frac{1}{2}(\eta_1(t)\pm\eta_2(t)),
\EEA
and then solved directly:
\BEA
\label{boa}
r_\pm(t)=e^{-(a\pm g)t}r_\pm(0)+\frac{1}{2}\int_0^t\d s\,e^{-(a\pm g)s}
(\eta_1(t-s)\pm \eta_2(t-s)).
\EEA

Eqs.~(\ref{babek}, \ref{zarzand}) 
produce for the coarse-grained velocity difference
(osmotic velocity) of the first and the second particle, respectively,
\BEA
\label{bakht}
u_1(\xb,t)=\frac{T\,(\sigma_{22}x_1-\sigma_{12}x_2)}
{\sigma_{11}\sigma_{22}-\sigma^2_{12}},\qquad
u_2(\xb,t)=\frac{T\,(\sigma_{11}x_2-\sigma_{12}x_1)}
{\sigma_{11}\sigma_{22}-\sigma^2_{12}}.
\EEA

The sufficient condition (\ref{dodo1}) reduces to
\BEA
\langle\,(\,\Delta u_1+\zeta \Delta u_2\,)^2\,\rangle+
\langle\,(\,\Delta x_1+\epsilon \Delta x_2\,)^2\,\rangle=
\frac{T^2\,(\sigma_{22}+\sigma_{11}- 2\zeta\sigma_{12})}
{\sigma_{11}\sigma_{22}-\sigma^2_{12}}+
\sigma_{22}+\sigma_{11}+ 2\epsilon\sigma_{12}
<4T.
\label{dard}
\EEA
A simpler expression is gotten for 
$\sigma_{11}=\sigma_{22}$~
\footnote{This equality will be satisfied for two identical brownian
particles if their initial conditions are the same.}:
\BEA
\label{burunduk}
\frac{T^2}{\sigma_{11}+\zeta\sigma_{12}}+\sigma_{11}
+\epsilon\sigma_{12}<2T.
\EEA

This condition is naturally not satisfied for $\sigma_{12}=0$, as the
minimal value of $\sigma_{11}+T^2/\sigma_{11}$ over $\sigma_{11}$ is
just equal to $2T$. For the correlated situation $\sigma_{12}>0$ and
we apply (\ref{burunduk}) for $\zeta=1$, $\epsilon=-1$, while for the
anti-correlated situation $\sigma_{12}<0$ we use in the same way
(\ref{burunduk}) for $\zeta=-1$, $\epsilon=1$. With these we get from
(\ref{burunduk}) the following condition
\BEA
(\sigma_{11}-T)^2<\sigma_{12}^2+2T|\sigma_{12}|.
\label{konkord}
\EEA
It is seen that this sufficient condition for the brownian
entanglement is symmetric with respect to correlation and
anticorrelation, and it can be satisfied if $\sigma_{11}$ is
sufficiently close to $T$ and $|\sigma_{12}|$ is finite.
Eq.~(\ref{konkord}) can be also satisfied when the particles are
either correlated or anti-correlated sufficiently strongly, e.g., when
$|\sigma_{12}|$ is sufficiently large and sufficiently close to its
upper bound $\sigma_{11}$. Note that (\ref{konkord}) is not satisfied
for $T\to 0$. This is natural, since there are no brownian uncertainty
relations in this limit.

Thus the brownian entanglement can exist for sufficiently strongly
fluctuating and/or sufficiently strongly interacting brownian particles. 

Let us check (\ref{burunduk}) more specifically with the stationary
(equilibrium) state of the brownian particles
which is established for long times provided the potential energy 
$U(x_1,x_2)$ is (strictly) positive, and where 
\BEA
\sigma_{11}=\sigma_{22}=\frac{Ta}{a^2-g^2},\qquad
\sigma_{12}=-\frac{Tg}{a^2-g^2}.
\EEA
These relations can be obtained either by directly solving (\ref{boa})
and then taking the limit $t\to \infty$, or directly via the stationary
distribution for the particles which is known to be Gibbsian: 
$P(x_1,x_2)\propto \exp[-U(x_1,x_2)/T]$.

Eq.~(\ref{konkord}) 
reduces to \footnote{Conditions (\ref{ka1}) are compatible with the stability
requirement $a>|g|$, which comes from demanding positivity of $U(x_1,x_2)$.
Indeed, once $|g|$ satisfies (\ref{ka1}), $a>|g|$
becomes $a(2+a)>(a-1)^2$ and is satisfied provided $a$ is not very small. }
\BEA
\label{ka1}
|g|>-1+\sqrt{1+(a-1)^2}.
\EEA
Thus if $|g|$ is large enough, 
there can be entanglement in the equilibrium state.

\subsection{Entanglement for two non-interacting brownian particles.}

In the quantum case two subsystems can be entangled even if they
interacted in the past, and they do not interact at the moment when
the entanglement is tested. This is also the case with the brownian
particles, as we show now. 

Return to
the above system of two harmonically interacting brownian
particles and assume that they did interact for $t<0$ but are not
interacting for positive times\footnote{
To avoid possible misunderstanding, recall that
the thermal baths acting on two particles were taken to be completely
independent of each other, so that there is no influence
via them. }. One has from (\ref{lacot})
(or from (\ref{boa})~):
\BEA
\sigma_{jj}(t)=e^{-2at}\sigma_{jj}(0)+\frac{T}{a}(1-e^{-2at}),\qquad
\sigma_{12}(t)=e^{-2at}\sigma_{12}(0),
\qquad j=1,2.
\EEA

We assume that the condition (\ref{burunduk}) (with
$\sigma_{11}=\sigma_{22}$ at all times) was satisfied at the initial
time, so that the brownian entanglement was present.  For positive
times and $a>0$, $\sigma_{12}(t)$ gradually disappears. This means
that for long times the sufficient condition (\ref{burunduk}) for the
brownian entanglement will not be valid.  Clearly, no brownian
entanglement can persists for very long times, since the particles
become in this limit completely non-correlated, i.e., the common
probaility distribution for $t\to\infty$ factorizes
$P(x_1,x_2,t)=P(x_1,t)P(x_2,t)$ (compare with our discussion after
(\ref{babek})).

It is however to be stressed that for certain not very long times we
shall surely have the condition (\ref{burunduk}) satisfied, at least
once it was satisfied initially.  Moreover, if even this condition was
not satisfied at $t=0$, it can curiously get valid for some (not very
long) times.  Here is an example.  For free brownian particles with
$a=0$, the correlation $\sigma_{12}(t)$ does not change in time at all,
but instead the dispersion of the coordinates increases linearly with
time: $\sigma_{jj}(t)=\sigma_{jj}(0)+2Tt$.  Condition (\ref{konkord})
for the presence of entanglement now reads:
\BEA
\label{geliogabal}
|\sigma_{11}(0)-T+2Tt|<\sqrt{\sigma_{12}^2(0)+2T|\sigma_{12}(0)|}.
\EEA
Even if this condition was not valid at $t=0$,
$|\sigma_{11}(0)-T|>\sqrt{\sigma_{12}^2(0)+2T|\sigma_{12}(0)|}$,
and additionally $\sigma_{11}(0)<T$, Eq.~(\ref{geliogabal})
can still be satisfied for 
\BEA
t_-<t<t_+,
\EEA
where
\BEA
t_\pm=\frac{T-\sigma_{11}(0)\pm\sqrt{\sigma^2_{12}+2T|\sigma_{12}|}}{2T}.
\EEA

Our general conclusion is that the brownian entanglement is
possible if the two brownian particles do interact or had interacted
strongly enough. The absence of interactions at present need not always
to destroy the brownian entanglement, for some finite times it may even
facilitate the sufficient conditions for its existence, as we saw above.

\section{Local osmotic velocities.}
\label{artashir}

In section \ref{III} we have seen that the definition of the coarse-grained
velocities (\ref{dish1}, \ref{dish2}, \ref{babek}) is given 
via the ensemble $\Sigma(x_1,x_2,t)$
in the common context of the two brownian particles. 

Let us now turn to the local velocities determined
with respect to the ensemble $\Sigma_1(x_1,t)$, which is obtained
by measuring at time $t$ only the coordinate $x_1$ of the first particle.

There are two completely equivalent ways for
determining the coarse-grained velocities over this ensemble. The
first way amounts to repeating definitions (\ref{dish1}, \ref{dish2})
for the ensemble $\Sigma_1(x_1,t)$
\BEA
\nu_{\pm,1}(x_1,t)=\pm \,{\rm lim}_{\ep\to +0}
\,\int\d y_1\,\frac{y_1-x_1}{\ep}\,P(y_1,t\pm \ep|x_1,t),
\label{papakan}
\EEA
where all the involved probability distributions contain no references
on the second brownian particle. The corresponding osmotic
velocity reads in complete analogy to (\ref{babek}):
\BEA
\label{ga1}
\mu_1(x_1,t)=\nu_{-,1}(x_1,t)-\nu_{+,1}(x_1,t).
\EEA

The second way is to note that if observer 1 has
no information at all from observer 2, then he effectively sums the ensemble
$\Sigma(x_1,x_2,t)$ over all possible results of 
the second coordinate $x_2$ at time $t$ with
his result $x_1$ being fixed. Each value of $x_2$ during this summation
is then met with probability $P(x_2,t|x_1,t)$, and this results in
\BEA
\label{ga2}
\nu_{\pm,1}(x_1,t)=\int\d x_2\,v_{\pm,1}(x_1,x_2,t)\,P(x_2,t|x_1,t).
\EEA
The equivalence between definitions (\ref{papakan}) and (\ref{ga2})
can be established via (\ref{shun}, \ref{katu}) and the Bayes formula
\footnote{Note that for obtaining the local 
coarse-grained velocities $\nu_{\pm,1}(x_1,t)$ we had to average
$v_{\pm,1}(x_1,x_2,t)$ over the conditional distribution
$P(x_2,t|x_1,t)$, and {\it not} over the unconstrained probability of
the second coordinate $P(x_2,t)$. This in-equivalence can be illustrated
with help of the following fact of probability theory concerning three
random variables $A,B,C$:
$\sum_CP(C)\,P(A|BC)\not=\sum_CP(C|B)\,P(A|BC)=P(A|B)$.  
Would we adopt the second possibility, e.g.,
$\tilde{u}(x_1,t)=
\int\d x_2\,u(x_1,x_2,t)\,P(x_2,t)$, this
would lead us to an explicitly common-context (non-local) quantity,
e.g., for the above example of harmonic oscillators we would get
$\tilde{u}(x_1,t)=\frac{T\sigma_{22}(t)x_1}
{\sigma_{11}(t)\sigma_{22}(t)-\sigma^2_{12}(t)}$, which explicitly
depends on the dispersion $\sigma_{22}(t)$ of the second particle. If
the two particles do not interact for positive times, but did interact
in the past (for $t<0$), $\tilde{u}(x_1,t)$ can still be influenced,
e.g., by external fields which apply on the second particle for
positive times (apparent or false non-locality).  }.

Completely similar definitions can be given for the second particle,
employing the ensemble $\Sigma(x_2)$ obtained by measuring the second
particle's coordinate only.  For the corresponding osmotic component one
now has from (\ref{ga1}, \ref{ga2}, \ref{babek}):
\BEA
\mu_j(x_j,t)
=-T_j\partial_{x_j}P(x_j,t), \qquad j=1,2.
\EEA

For the brownian particle in the harmonic potential, as described 
e.g. by (\ref{taratuj}), one has
\BEA
\label{barbos}
\mu_j(x_j,t)=T_j\frac{x_j}{\sigma_{jj}(t)}.
\EEA

If now the definition (\ref{tartar}) of non-entanglement is applied to the
common distribution function
\BEA
P(x_1,\mu_1,x_2,\mu_2)=
P_1(x_1)\,\delta(\,\mu_1-\mu_1(x_1)\,)\,
P_2(x_2)\,\delta(\,\mu_2-\mu_2(x_2)\,),
\EEA
then it is seen to be satisfied trivially: No entanglement occurs
with locally defined osmotic velocities

\section{Discussion.}
\label{samson}

Let us compare in more detail the physical meaning of the quantum
mechanical entanglement versus its brownian analog.

\begin{itemize}

\item
In analogy to continuous-variable quantum entanglement, brownian
entanglement is defined as a type of correlation between the
coordinates and the changes of the coarse-grained velocities (osmotic
velocities) of two brownian particles, which is impossible to
reproduce by mixing non-correlated |that is, referring to each particle
separately| distributions.

\item The brownian uncertainty relation (\ref{gogo}) between the
coordinate and the change of the coarse-grained velocity is necessary
for the very existence of the brownian entanglement.  This is again
similar to the quantum situation, where the analogous role of a
necessary condition is being played by the quantum uncertainty
relations.  

\item
For checking quantum entanglement via sufficient condition (\ref{klm})
one needs to make measurements of non-commuting coordinate $\hat{x}_k$
and momentum $\hat{p}_k$ for each quantum particle ($k=1,2$). To this
end one has to have two different ensembles of the particle $\1$ and
$\2$ (compare with our discussion in section \ref{oper}). In contrast,
the brownian entanglement involves only consecutive coordinate
measurements done on a single ensemble of the brownian pairs.

\item
The momentum operator $\hat{p}$ in quantum mechanics |though being
equal in Heisenberg representation to the time-derivative of the
coordinate operator $\hat{p} =m\frac{\d}{\d t}\hat{x}$|
does not in general fully characterize the intuitive notion of ``change of the
coordinate for an infinitesimal time''
~\footnote{\label{get} This is a general
point. A difference $\hat{A}(t)-\hat{A}(0) =\int_0^t\d t\frac{\d}{\d
t}\hat{A}$ of Heisenberg operators does not fully
characterize the change in
time of the observable $\hat{A}$, because there are ensembles
described by (time-independent in the Heisenberg representation)
states $|\psi\rangle\langle\psi|$ for which $(\hat{A}(t)-\hat{A}(0))
|\psi\rangle=0$. This seems to imply that the value of $\hat{A}$ did
not change at all, but this is not correct, since the above eigenvalue
relation may be still compatible |due to non-commutativity
$[\hat{A}(t),\hat{A}(0)]\not =0$| with different statistics of
$\hat{A}(t)$ and $\hat{A}(0)$, e.g.,
$\langle\psi|\hat{A}^3(t)|\psi\rangle
\not =\langle\psi|\hat{A}^3(0)|\psi\rangle$. For more elaborated
discussion and concrete examples see \cite{work}. For unbound
operators, such $\hat{x}$ and $\hat{p}$, the above reasonings may need
to be technically modified, since eigenstates of unbound
operators are not normalizable. In this context it may suffice to require
$(\hat{A}(t)-\hat{A}(0))|\psi\rangle\approx 0 $.  }.

In contrast, the brownian entanglement is about the coordinates and
the coarse-grained velocities which do characterize the change of the
coordinates on the coarse-grained scale of time.

\item Both quantum entanglement and its brownian analog can exist for
subsystems which interacted in the past, but do not interact at the
present.

\item In quantum mechanics the sufficient condition 
(\ref{klm}) for entanglement involves correlations between the
coordinate operators $\hat{x}_1$ and $\hat{x}_2$, and the momentum
operators $\hat{p}_1$, $\hat{p}_2$ of the subsystems $\1$ and $\2$.
Quantum entanglement can only be found after the results got via
measurements on the corresponding subensembles for the subsystems $\1$
and $\2$ are put together.  The situation with the brownian
entanglement is similar, since it also requires correlation
experiments.

\item Here is finally
the main conceptual difference between the quantum
entanglement and its brownian analog.  In quantum mechanics the above
operators of coordinate and momentum pertain to the corresponding
subsystems $\1$ and $\2$ {\it independently} of the full system context.  By
this we mean that all the statistics of, e.g., $\hat{p}_1$ can be
collected via local measurements on the corresponding quantum
subensemble, whether or not this subensemble forms a
part of any larger ensemble.

In contrast, the very definition of the coarse-grained velocities
(\ref{dish1}, \ref{dish2}, \ref{babek}) involves a global (that is,
depending on the two subsystems) ensemble $\Sigma(\xb,t)$. As seen in
section \ref{artashir}, the purely local definition of coarse-grained
velocities can also be given, but there will not be any entanglement
on that level, for the same reason as there is no entanglement in
other classical systems (see section
\ref{claa}).

This conclusion on the main difference is close to the analogous
conclusion of Ref.~\cite{spr}, which discusses similarities between
quantum entanglement and certain correlations in classical optics.

\end{itemize}

\section{Possibilities of experimental realization.}
\label{expo}

One hopes that an experimental verification of the brownian
entanglement is going to be easier than that of its quantum analog:
since Brown's discovery in 1828, various examples of brownian motion
are routinely observed in many systems. On the other hand, the basic
conditions needed for observation of the brownian entanglement amount
to two coupled brownian particles and a resolution of the brownian
motion sufficient for observing the osmotic velocities.  Recall that
on the coarse-grained time scale the osmotic velocities can be
visualized as average kicks got by the brownian particle due to its
interaction with bath particles.

Here we discuss two experimentally studied examples of brownian
motion. The pecularity of these examples is that the brownian motion
can be detected with human eyes (without microscopes).  These are thus
examples of macroscopic brownian motion. Recall that the typical
examples of brownian motion involve much smaller scales: pollen
molecules in water observed originally by Brown had a size $\approx
10^{-3}$ cm, and even smaller lengths are typical for many other
realizations of the brownian motion.

The first experiment \cite{e1} considers a two-dimensional circular
container with elastic walls. Inside of the container there are
motorized balls with a mass of 120 g and a diameter of 8 cm. Each ball
is driven by a battery-powered motor inside, and moves chaotically due
to elastic collisions with the walls of the container and with other
motorized balls. The set of balls (with concentration 50 balls/m$^2$)
models the bath particles. The brownian particles are modelled with a
set of ping-pong balls (with a mass 2 g and diameter 4 cm) which
undergo random collisions with the motorized balls.  The primary
purpose of \cite{e1} was to study polymer statistics, so that the
ping-pong balls were jointed into a long chain by means of
springs. The resulting coupled brownian motion produced a probability
distribution for the end-to-end distance of the ping-pong chain in a
close agreement with the existing theories of two-dimensional
(self-avoiding) random walk.  In this experiment Langevin equation as
such was not tested directly, but there were several indirect reasons
supporting its validity \cite{e1}. This system would fit for observing
the brownian entanglement, because almost all needed ingredients of
this phenomenon are present.

A curious pedagogical analog of the above experiment is the motion of
a child toy called Bumble Ball \cite{e0}: a plastic sphere about 11 cm
in diameter with a symmetrical pattern of rubber knobs extending about
3 cm from its surface. An internal motor rotates an off-axis mass such
that when activated and placed on a hard surface, a Bumble Ball
simulates a random walk \cite{e0}. Joining two such balls with a
spring and placing them into a container creates a toy model which
might provide at least the most rough indications of the brownian
entanglement.

The second experiment \cite{e2} studied a single sphere (a ping-pong
ball of a mass 2.5 g and diameter 4 cm) rolling stochastically in an
upflow of gas under flow speed 280 cm/c. The sphere rolls
stochastically due to the turbulence it generates in the gas stream
(the corresponding Reynolds number is $\approx$40). The study focussed
on the full time-dependent dynamics of the sphere, as captured by
video imaging.  It appears that the dynamics is that of the
underdamped brownian particle in a harmonic potential. This
correspondence was thoroughly checked from various perspectives. When
looking for the brownian entanglement in such a situation, it would be
necessary to go to the overdamped motion regime (e.g., by decreasing
the mass of the sphere and by increasing the flow speed of the stream)
and to add there a second sphere.

\section{Conclusion.}

We have uncovered the phenomenon of brownian entanglement: a
correlation effect between the coordinates and the coarse-grained
velocities of two classical brownian particles, which resembles the
quantum entanglement. In contrast to the latter, which is presently
given a fundamental status, the brownian entanglement |as the very
subject of statistical physics| arises out of coarse-graining
(incomplete description) reasons. In that respect it is similar to
other basic relations of the statistical physics, such as the second
law \cite{work}.  In the present situation the coarse-graining comes
due to the time-scale separation: the evolution of the momenta of the
brownian particles is very fast and cannot be resolved on the
time-scales available to the experiment. The idea of time-scale
separations is one of the most pertinent ones in non-equilibrium
statistical physics. In a qualitative form it appears already in good
textbooks on this subject
\cite{landau,ma}, and has been since then formalized in various
contexts and on various levels of generality
\cite{landauer,kurchan,theo,at,shu}.

Once there is time-scale separation between the (stochastic) motion of
the coordinate and the momentum of the brownian particles, the
operational definition of velocity via the rate of the coordinate
leads to the coarse-grained velocity which is not equal to the real
momentum. The change of the coarse-grained velocity is controlled by
the brownian uncertainty relation. Moreover, the coarse-grained
velocities appear to be contextual random quantities, i.e., they
depend on the concrete setting of the coordinate measurement used to
define them.  These two aspects (uncertainty relations and
contextuality) suffice to define brownian entanglement, similarly to
quantum entanglement, as impossibility to prepare the common
distribution of the coordinates and the coarse-grained velocities of
two brownian particles via mixing locally independent (non-correlated)
distributions referring to the two particles separately. Again in
analogy to quantum entanglement, brownian entanglement can be
witnessed via the uncertainty relations.

In this paper we demonstrated the brownian entanglement on the simplest,
exactly solvable, one-dimensional models of two interacting brownian
particles. It should be however kept in mind that interacting brownian
particles (random walks) are basic for several fields of modern
statistical physics, such as colloids \cite{collo} or polymers
\cite{gros}. One of the most transparent examples from the polymer
science is a DNA macromolecule which consists of two interacting random
walks (strands) \cite{gros}.  As we argued, the experimental observation
of the brownian entanglement might be easier than those of the quantum
entanglement, since there are realizations of macroscopic brownian
motion that are visible with the human eye. 

\acknowledgments 
A.E. A thanks E. Mamasakhlisov for useful discussions.
The work of A.E. A is part of the research programme of the Stichting voor 
Fundamenteel Onderzoek der Materie (FOM, financially supported by 
the Nederlandse Organisatie voor Wetenschappelijk Onderzoek (NWO)).

\appendix

\section{}

This Appendix has two closely related purposes. First, we shall apply
the definition (\ref{dish1}, \ref{dish2}) of the coarse-grained
velocities in a non-overdamped situation, that is, we shall take
$\eps$ in those definitions much smaller than the characteristic
relaxation time of the (real) momenta. We shall convince ourselves
that the expected answer is obtained, relating the coarse-grained
velocities to the average (real) momentum. In addition, we shall see
that the the coarse-grained velocity difference (osmotic velocity)
(\ref{babek}) disappears in this situation, i.e., the definitions
(\ref{dish1}, \ref{dish2}) become equivalent. Second, based on an
exactly solvable situation, we shall follow in detail to the behavior
of the coarse-grained velocity as a function of $\eps$. For simplicity
we operate only with one brownian particle.

\subsection{}

The evolution of the common probability distribution 
of the coordinate
$x$ and the momentum $p$ of the brownian particle is described by the
following Fokker-Planck-Kramers-Klein equation \cite{risken}
\BEA
\partial_t{\cal P}=-\frac{p}{m}\partial _x{\cal P}+\partial_p
(\frac{\gamma}{m}p+V'(x)+\gamma T\partial_p){\cal P},
\label{t1}
\EEA
where $m$, $\gamma$ and $V(x)$ are mass, damping constant and potential, 
respectively, and where
\BEA
\label{t2}
{\cal P}(x,p,t|x',p',t'),\qquad t\geq t',\qquad
{\cal P}(x,p,t|x',p',t)=\delta(x-x')\,\delta(p-p'),
\EEA
is the conditional probability to move from $(x',p')$ at time $t'$ 
to  $(x,p)$ at time $t$. 
This equation corresponds to the Langevin equation
\BEA
m\ddot{x}=-ax-\gamma\dot{x}+\eta(t),\qquad
\langle\eta(t)\rangle=0,\qquad
\langle\eta(t)\eta(t')\rangle=2\gamma T\delta(t-t'),
\label{t5}
\EEA
where $-\gamma\dot{x}$ is the friction force, and where $\eta(t)$
is the random gaussian white noise.

Applying the definition (\ref{dish1}) of the coarse-grained velocity
and using (\ref{t1},\ref{t2}) one has
\BEA
\label{dish101}
\nu_{+}(x,t)&&={\rm lim}_{\epsil\to +0}
\,\int\d y\,\frac{y-x}{\epsil}\,{\cal P}(y,t+\ep|x,t)\nonumber
\\
&&={\rm lim}_{\epsil\to +0}
\,\int\d y\,\d p\,\d p'\,
\frac{y-x}{\epsil}\,{\cal P}(y,p,t+\ep|x,p',t)\,
{\cal P}(p',t|x,t)\nonumber\\
&&=-\int\d y\,\d p\,(y-x)\,{\cal P}(p,t|x,t)\,
\frac{p}{m}\,\partial_y\delta(x-y)\nonumber\\
&&=\int\d p\,{\cal P}(p,t|x,t)\,
\frac{p}{m},
\EEA
where ${\cal P}(p,t|x,t)$ is the conditional probability for having the
momentum equal to $p$ at time $t$, provided the coordinate was equal to 
$x$ at the same time.

Likewise,

\BEA
\nu_{-}(x,t)&&={\rm lim}_{\epsil\to +0}
\,\int\d y\,\frac{x-y}{\ep}\,P(y,t-\ep|x,t)\nonumber\\
&&={\rm lim}_{\epsil\to +0}
\,\int\d y\,\d p\,\d p'\,
\frac{x-y}{\epsil}\,
{\cal P}(x,p,t|y,p',t-\ep)\,
\frac{  {\cal P}(y,p',t-\ep)}{{\cal P}(x,t)}\,
\nonumber\\
&&=\int\d y\,\d p\,(y-x)\,
\frac{ {\cal P} (y,p,t) }{ {\cal P}(x,t) }\,
\frac{p}{m}\,\partial_x\delta(x-y)\nonumber\\
&&=\int\d p\,{\cal P}(p,t|x,t)\,
\frac{p}{m}=\nu_{+}(x,t).
\label{dish200}
\EEA

The fact that $\nu_{-}(x,t)=\nu_{+}(x,t)$ can be easily generalized 
to any number of interacting brownian particles. 

Now if there are no correlations between the momentum and coordinate
${\cal P}(p,t|x,t)={\cal P}(p,t)$, then $\nu_{-}(x,t)=\nu_{+}(x,t)$
reduce to the usual average momentum.  This is the case, in
particular, for the deterministic situation, where ${\cal
P}(x,p,t)=\delta(x-x(t)\,)\,\delta(p-p(t)\,)$.

Thus provided $\eps$ has been taken much smaller than any relevant
time-scale related to the coordinate and/or the momentum, the
coarse-grained velocity reduces to the average momentum as it should.
In this situation the osmotic velocity is zero.

\subsection{}

Now let us show on the exactly solvable situation with a harmonic
potential
\BEA
V(x)=\frac{ax^2}{2},
\EEA
that if the characteristic 
relaxation times of the momentum 
\BEA
\tau_p=\frac{m}{\gamma},
\EEA
and the
coordinate 
\BEA
\tau_x=\frac{\gamma}{a},
\EEA
are well separated: 
\BEA
\tau_p\ll\tau_x, 
\EEA
and if additionally
\BEA
\tau_x\gg
\epsil \gg
\tau_p,
\label{dada}
\EEA
one gets back the values for osmotic velocity which were obtained
in the main text by means of the overdamped Fokker-Planck equation
(\ref{d1}).

The case with $V(x)=ax^2/2$ can be solved either directly from 
(\ref{t1}), or using the equivalent Langevin equation (\ref{t5}).
The solution of the latter
|obtained, e.g., via Laplace transformation| reads:
\BEA
\label{t7}
x(t)=x(0)g(t)
+\frac{1}{m}p(0)f(t)+\frac{1}{m}\int_0^{t}\d t'
f(t-t')\eta (t'),
\EEA
where 
\BEA
\label{t77}
f(t)=\frac{e^{-\omega_2t}-e^{-\omega_1t}}{\omega_1-\omega_2},
\qquad
g(t)=\frac{\omega_1e^{-\omega_2t}-
\omega_2e^{-\omega_1t}}{\omega_1-\omega_2},
\EEA
\BEA
\label{brut1}
\omega _{1,2}=\frac{\gamma}{2m}
\left (1\pm \sqrt{1-\frac{4am}{\gamma ^2}}\,\, \right ).
\EEA
For simplicity reasons we shall put $x(0)=p(0)=0$. This can be done, since
these quantities are assumed to be independent of the noise by the very
definition of the considered stochastic process. 

As the process (\ref{t7}) is gaussian, the 
two-time probability distribution of the coordinate
can be written down as:
\BEA
{\cal P}(y,s;x,t)=\frac{\sqrt{d}}{2\pi}
\exp-\frac{1}{2}\left[
A_{11}y^2+A_{22}x^2+2A_{12}xy
\right],
\EEA
where
\begin{gather}
d\equiv\sigma(t,t)\sigma(s,s)-\sigma^2(s,t),\\
\left (\begin{array}{rr}
A_{11} & A_{12}\\
A_{12} & A_{22}
\end{array}\right )^{-1}=
\left (\begin{array}{rr}
\sigma(s,s) & \sigma(s,t)\\
\sigma(s,t) & \sigma(t,t)
\end{array}\right )=
\frac{1}{d}
\left (\begin{array}{rr}
\sigma(t,t) & -\sigma(s,t)\\
-\sigma(s,t) & \sigma(s,s)
\end{array}\right ),
\end{gather}
and where $\sigma(s,t)$ is the correlation function of the coordinate,
\BEA
\sigma(s,t)
=\langle x(s)x(t)\rangle
=\frac{2\gamma T}{m^2}\int_0^s\int_0^t\d t_1\d t_2\,
f(s-t_2)f(t-t_2)\delta(t_1-t_2)
=\frac{2\gamma T}{m^2}\int_0^{{\rm min}(s,t)}\d t'
f(t')f(t'+|s-t|).
\label{t10}
\EEA

Using:
\BEA
\int\d y\,y\,{\cal P}(y,s|x,t)=
\int\d y\,y\,\frac{{\cal P}(y,s;x,t)}{{\cal P}(x,t)}=
-x\,\frac{A_{12}}{A_{11}}=x\,\frac{\sigma(s,t)}{\sigma(t,t)},
\EEA
one gets keeping $\ep$ finite and fixed
\BEA
\label{mob1}
\nu_{+}(x,t)&&=\,\int\d y\,\frac{y-x}{\epsil}\,{\cal P}(y,t+\ep|x,t)
=\frac{x}{\epsil}\,
\left(\frac{\sigma(t+\ep,t)}{\sigma(t,t)}-1\right),\\
\nu_{-}(x,t)&&=
\,\int\d y\,\frac{x-y}{\ep}\,P(y,t-\ep|x,t)
=\frac{x}{\epsil}\,
\left(1-\frac{\sigma(t-\ep,t)}{\sigma(t,t)}\right),
\label{mob2}
\EEA
for any $\epsil\geq 0$ smaller than $t$: $t-\epsil>0$.

If now $\ep$ is smallest time-scale,
\BEA
\nu_{-}(x,t)-\nu_{+}(x,t)
=\frac{2\gamma T}{m^2}\,\frac{x\ep}{\sigma(t,t)}
\left[
\int_0^t\d t'\,\dot{f}^2(t')-f(t)\dot{f}(t)
\right]+{\cal O}(\ep^2)
\EEA
goes to zero with $\epsil\to 0$ as was predicted above.

As seen from (\ref{brut1}), the overdamped limit is realized with the
dimensionless parameter $4am/\gamma^2$ being small:
\BEA
\frac{4am}{\gamma^2}\ll 1,
\EEA
and then the charactersitc relaxation times of the momentum
and the coordinate [obtained from (\ref{t7}, \ref{t77}, \ref{brut1})]
are well-separated
\BEA
\label{esher}
\tau_p=\frac{m}{\gamma}\,\approx \,
\frac{1}{\om_1}\,\ll\, \tau_x=\frac{\gamma}{a}
\,\approx \,\frac{1}{\om_2}.
\EEA
Now if 
\begin{gather}
\label{dd}
\omega_1 t\gg 1,\qquad
\omega_1 s\gg 1,\qquad
\omega_1|s-t|\gg 1,
\end{gather}
one gets from (\ref{t10})
\BEA
\label{kaban}
\sigma(s,t)=
\frac{T}{a}\left(e^{-\omega_2|s-t|}-e^{-\omega_2(t+s)}
\right),
\EEA
which coincides with the corresponding correlator obtained from the
overdamped Fokker-Planck equation directly. Taking now $s=t\pm \ep$ in
(\ref{mob1}, \ref{mob2}) and employing (\ref{kaban}) with the limit
$\ep\to 0$ |which now should be understood in the context of conditions
(\ref{esher}, \ref{dd})| one can get the expressions for
$\nu_{\pm}(x,t)$ which can alternatively be derived from (\ref{shun},
\ref{katu}), if we put there $N=1$ (only one brownian particle),
$f(x)=-ax/\gamma$ (the linear force divided over the friction
constant),
and $P(x,t)\propto \exp[-\frac{x^2}{2\sigma(t,t)}]$ (gaussian
distribution
function).

\end{document}